# Experimental confirmation of the delayed Ni demagnetization in FeNi alloy

S. Jana[1, *], R. Knut[2], S. Muralidhar[3], R. S. Malik[2], R. Stefanuik[2], J. Åkerman[3], O. Karis[2], C. Schüßler-Langeheine[1], N. Pontius[1]

[1]Helmholtz-Zentrum Berlin für Materialien und Energie GmbH, Albert Einstein Straße 15, 12489 Berlin, Germany

[2]Department of Physics and Astronomy, Uppsala University, Box 516, 75120 Uppsala, Sweden

[3]Department of Physics, University of Gothenburg, 412 96 Gothenburg, Sweden

Abstract:

Element-selective techniques are central for the understanding of ultrafast spin dynamics in multi-element materials like magnetic alloys. Recently, though, it turned out that the commonly used technique of transverse magneto-optical Kerr effect (T-MOKE) in the EUV range may have linearity issues including unwanted cross talk between different elemental signals. This problem can be sizeable, which puts recent observations of ultrafast spin transfer from Fe to Ni sites in FeNi alloys into question. In this study, we investigate the Fe-to-Ni spin transfer in a cross-talk-free time-resolved X-ray magnetic circular dichroism (XMCD) experiment with a reliable time reference. We find a very similar Fe and Ni dynamics with XMCD as with T-MOKE from identical samples. Considering the non-linearities of the T-MOKE response, the agreement with our findings appears fortuitous. We discuss possible reasons why T-MOKE seems to give accurate results in this case. Our data provide the ongoing discussion about ultrafast spin-transfer mechanisms in FeNi systems with a sound experimental basis.

Introduction:

Over last two decades, element resolved studies of magnetization dynamics became possible with the development of short-pulsed EUV (extreme ultra violet) and X-ray sources. They allowed probing various intrinsic quantities, such as inter-atomic exchange interaction, inter-atomic spin transfer, spin and orbital contributions, as well as layer resolved dynamics. [1,2,11–15,3–10] The most commonly used techniques in this field are transverse magneto-optical Kerr effect (T-MOKE) at EUV energies [9–12] and X-ray magnetic circular dichroism (XMCD) in the soft x-ray rang [13,14,16,17], probing element resolved magnetization dynamics via excitations from shallow and deep core levels, respectively.

Element-sensitive techniques are particularly important when different elemental species in one magnetic system show different dynamics [2,4–6,13,15]. One very interesting recent result in this context is the apparent delay of the demagnetization in Ni compared to that of the Fe in an FeNi alloy [2,6,11,12]: T-MOKE studies on these alloys find Ni to demagnetize after a considerable delay with respect to Fe. This puzzling observation motivated theoretical modelling and two different explanations have been put forward: Knut et al., explains the observed delay in Permalloy (Py) and $Cu_{0.4}Py_{0.6}$ sample adopting an inhomogeneous magnon generation scheme that provides the mechanism of the angular momentum transfer to the lattice. [5] This model uses a combination of electron-phonon and electron-magnon interaction, initially proposed by Hagg et al. [18]. On the other hand, Hofherr et al. [8] showed that optically induced spin transfer (OISTR) between the Fe and the Ni sublattice in FeNi alloy can explain the observed delay in Ni demagnetization as well. Both proposed mechanism are relevant beyond this particular effect as they describe fundamental processes in ultrafast spin manipulation in multi-element magnets.

---

[*]Author to whom correspondence should be addressed: sj.phys@gmail.com.

Remarkably, the experimental basis for these theories turned out rather unstable. The delay in the onset of demagnetization of Ni relative to that of Fe in $Fe_{1-x}Ni_x$ alloys has only been reported in experimental studies, that used T-MOKE at EUV [2,6,11,12]. In contrast to that, first XMCD data reported no delay for the Ni demagnetization in FeNi-alloy, but found substantially slower demagnetization for Fe. [15] This apparent discrepancy in the findings of the two experimental techniques is intriguing and disturbing; it might be caused by practical or fundamental limitations that both techniques have, but may as well have serious implications for the validity of the proposed mechanisms.

For 3d transition metals, T-MOKE is carried out at the $3p \rightarrow 3d$ ($M_{2,3}$) resonances in the EUV range and XMCD at the $2p \rightarrow 3d$ ($L_{2,3}$) resonances in the soft X-ray range. Both involve excitations into the magnetically relevant 3d states. While XMCD is directly probing the magnetic contribution to the X-ray absorption cross section, which quite rigorously scales with the magnetization [19–21], the relation between EUV T-MOKE signal and magnetization is somewhat more involved. In a recent work, some of us showed that the relation between the asymmetry measured in T-MOKE and sample magnetization can be non-linear even in equilibrium [22]. The response is rather complex including sign changes and depends on both, the photon energy of the probe pulse and the type of excitation of the spin system. So would a collinear change of the magnetic moment (Stoner-like excitation) in Fe result in a highly nonlinear response at probing energies above the $M_{2,3}$-edge resonance. Overall, such non-linearities may have affected the experimental observations discussed above.

Linearity may become even more of an issue when two resonances are probed that overlap in energy. The energy distance between the $M_{2,3}$ resonances of two neighboring 3d transition metals is smaller than the energy width of each single resonance in the T-MOKE signal, such that even the contributions from next-neighboring elements like Fe and Ni overlap. This may cause unwanted cross-talk between the dynamic signals from different elements. The effect can be a sizeable: The T-MOKE asymmetry of elemental Ni at the energy of the Fe-$M_{2,3}$ resonance is about half as large as the elemental Fe asymmetry at the same energy and of opposite sign [22]. One can easily imagine this to cause misleading experimental results. Furthermore, in dynamics studies with T-MOKE, effects may occur like energy shifts [22] or energy-dependent dynamic responses [23].

Cross-talk problems do not occur for the well-separated soft X-ray $L_{2,3}$ resonances; the energy difference between the Fe and Ni $L_{2,3}$ resonances exceeds the width of the elemental resonances by far. While the energy separation between resonances in the soft X-ray range safely excludes cross-talk, it causes other types of experimental problems in element-resolved dynamics studies. In the EUV, the close proximity of resonances allows for simultaneous detection of the different resonances, thus automatically guaranteeing a common time scale. In contrast to that, all sources of soft X-ray pulses available today cover only one resonance at a time. One therefore has to carefully reference time scales when studying effects like delayed dynamic response in one elemental subsystem.

Given this overall complexity, a referencing of one element-resolved techniques against the other appears a reasonable step towards providing a reliable experimental basis for theory. In this work, we directly compare the elemental magnetization dynamics in $Fe_{0.5}Ni_{0.5}$ utilizing both techniques for identical samples. Carefully avoiding all known sources of experimental problems, we can establish that the delay of the Ni demagnetization is indeed intrinsic to the sample properties and that – surprisingly – both techniques even quantitatively find very similar dynamical responses.

Experiments and results:

Time resolved T-MOKE was performed at the HELIOS Laboratory, Uppsala University, Sweden. [24] A near-infrared (NIR) pump pulse of wave length 800 nm (1.5 eV) and pulse length 35 fs, and EUV probe pulse of energies ranging between 40 – 70 eV and a pulse length of 20 fs are utilized in the T-MOKE-setup that measures the whole EUV spectra at once [12]. EUV photons are obtained via high harmonic generation with the same NIR pulse also used for pumping thus ensuring intrinsic synchronization. The measurement geometry and measurement protocol is described in detail in Ref. 4. As depicted in Fig. 1a, the sample was magnetized perpendicular to the scattering plane spanned by the directions of the incoming and reflected EUV photons; the polarization of the incoming EUV photons was linear and parallel to the scattering plane. In T-MOKE geometry, a change of the reflected intensity is observed upon reversal of the sample magnetization, which is utilized to probe the magnetization magnitude.

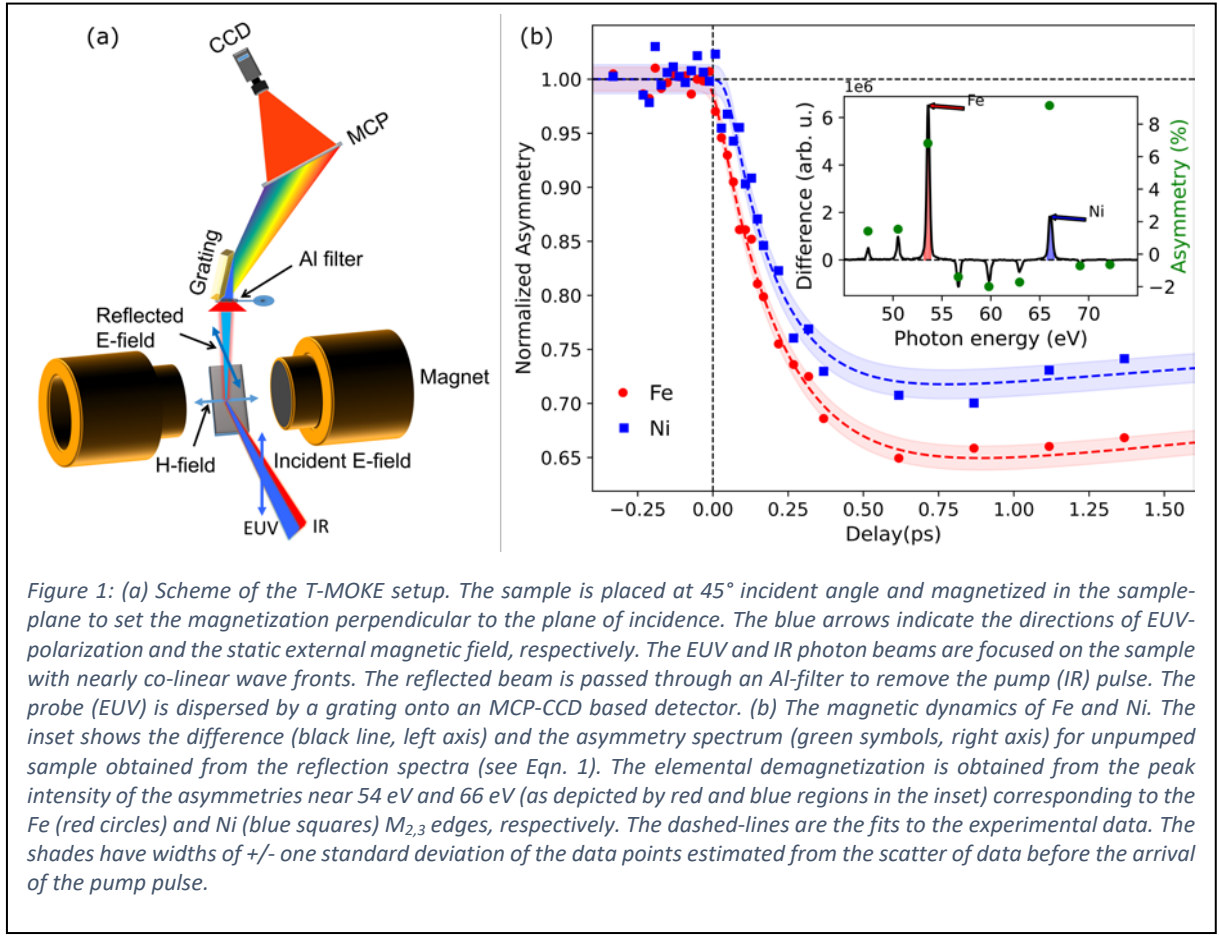

*Figure 1: (a) Scheme of the T-MOKE setup. The sample is placed at 45° incident angle and magnetized in the sample-plane to set the magnetization perpendicular to the plane of incidence. The blue arrows indicate the directions of EUV-polarization and the static external magnetic field, respectively. The EUV and IR photon beams are focused on the sample with nearly co-linear wave fronts. The reflected beam is passed through an Al-filter to remove the pump (IR) pulse. The probe (EUV) is dispersed by a grating onto an MCP-CCD based detector. (b) The magnetic dynamics of Fe and Ni. The inset shows the difference (black line, left axis) and the asymmetry spectrum (green symbols, right axis) for unpumped sample obtained from the reflection spectra (see Eqn. 1). The elemental demagnetization is obtained from the peak intensity of the asymmetries near 54 eV and 66 eV (as depicted by red and blue regions in the inset) corresponding to the Fe (red circles) and Ni (blue squares) $M_{2,3}$ edges, respectively. The dashed-lines are the fits to the experimental data. The shades have widths of +/- one standard deviation of the data points estimated from the scatter of data before the arrival of the pump pulse.*

The XMCD experiments were performed at the BESSY-II Femtoslicing facility (beamline UE56/1-ZPM and DynaMaX end-station) at Helmholtz-Zentrum Berlin [25]. Elliptically polarized X-ray pulses of 100 fs temporal width are produced via femto-slicing of electron bunches by means of an intense NIR pulse (of wavelength 800 nm and pulse width 50 fs) generated in a seeded amplifier. An NIR pulse produced in a separate amplifier, synchronized with the slicing pulse, is used as pump pulse for sample excitation. The repetition rate of the pump pulse is 3 kHz and that of the probe pulse is 6 kHz, such that alternatingly the signals with and without pump pulse (in the following referred to as pumped and unpumped signal, respectively) are detected. The measurement was performed in transmission geometry with the sample placed at 45° incidence angle with the X-ray beam. The magnetic field aligned parallel to the X-ray beam was inverted every 10 s to obtain the XMCD contrast. The X-ray photons transmitted through the sample were detected with an avalanche photo diode. [25]

Our sample is a 25 nm film of $Fe_{0.5}Ni_{0.5}$ alloy. It was deposited under identical deposition condition on Ta (3 nm)/Al (0.3 µm)/$Si_3N_4$ (0.2 µm) and on Ta (3nm)/Si (1 mm) by means of magnetron sputtering technique. The sample on the $Si_3N_4$ membrane was used to measure the XMCD in transmission; the Al layer acts as a heat sink to protect the sample from the laser-damage. The sample on the Si substrate was used to measure the T-MOKE signal in reflection. A Ta layer of 3 nm was deposited on top of the sample to protect against oxidation.

We start by discussing the T-MOKE results. The quantity considered in this experiment is the

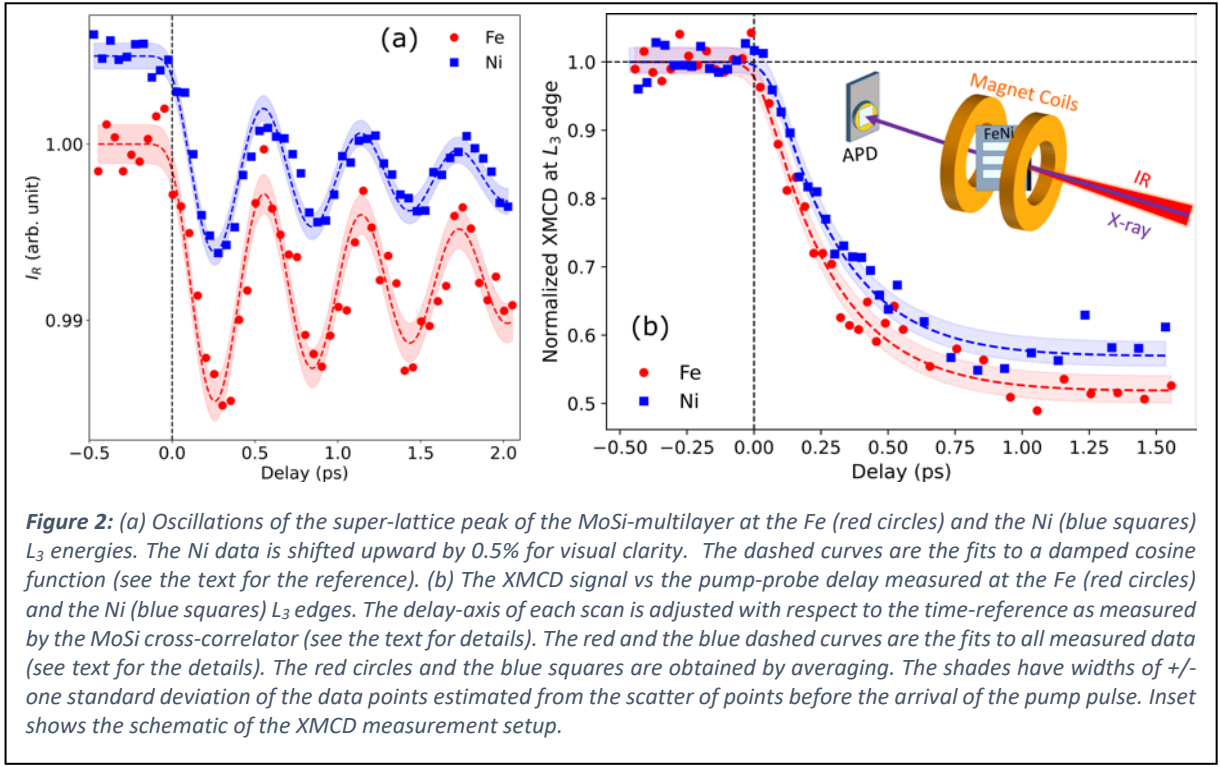

*Figure 2: (a) Oscillations of the super-lattice peak of the MoSi-multilayer at the Fe (red circles) and the Ni (blue squares) $L_3$ energies. The Ni data is shifted upward by 0.5% for visual clarity. The dashed curves are the fits to a damped cosine function (see the text for the reference). (b) The XMCD signal vs the pump-probe delay measured at the Fe (red circles) and the Ni (blue squares) $L_3$ edges. The delay-axis of each scan is adjusted with respect to the time-reference as measured by the MoSi cross-correlator (see the text for details). The red and the blue dashed curves are the fits to all measured data (see text for the details). The red circles and the blue squares are obtained by averaging. The shades have widths of +/- one standard deviation of the data points estimated from the scatter of points before the arrival of the pump pulse. Inset shows the schematic of the XMCD measurement setup.*

asymmetry spectrum $A(E)$, which is a function of the photon energy ($E$) as

$$A(E) = \frac{I_+(E) - I_-(E)}{I_+(E) + I_-(E)} \quad (1)$$

where $I_\pm(E)$ is the reflected intensity for opposite magnetization directions ($\pm M$). As pointed out in Ref. 22, the relation between asymmetry and magnetization in TMOKE is energy dependent. Here, we follow the standard protocol and probe the magnetization dynamics from measuring the asymmetry at energies that corresponds to the harmonics with peak asymmetries near the $M_{2,3}$ edges of Fe and Ni. The inset of Fig. 1b shows the difference and asymmetry spectrum, respectively. The strong peaks correspond to the $M_{2,3}$ edges of Fe (near 54 eV) and Ni (near 66 eV) as indicated by arrows. We note the complex energy dependence with a sign change in between the Fe and Ni resonance maxima.

To determine the element-resolved magnetic dynamics, asymmetry spectra were recorded with different pump-probe delays. The asymmetry at negative pump-probe delays was normalized at both, the Fe and the Ni $M_{2,3}$ edges; the resulting asymmetry curves are plotted against the pump-probe delay in the main panel of Fig. 1b which then represent the element resolved T-MOKE dynamics at these edges.

As the whole $A(E)$ spectrum is measured at each delay, the dynamics at Fe and Ni sites are probed simultaneously. A delay in the onset of Ni demagnetization to that of the Fe is evident from the data. A double exponential decay function fits both the curves featuring an exponential decay and a subsequent recovery:

$$m(t) = m_0 \left[1 - m_d \{1 - \exp(-(t-t_s)/t_d)\} \exp(-(t-t_s)/t_r)\right] \otimes G \quad (2)$$

Here $m(t)$ is the time-dependent magnetization, $m_0$ is the initial full magnetization magnitude. $m_d$ is the maximum portion of the quenched magnetization after excitation, $t_s$ defines the onset of the demagnetization, while $t_d$ and $t_r$ are the time constants of the decay and recovery, respectively. A convolution with a Gaussian, $G$, accounts for the experimental resolution. The results for the most relevant parameters are presented in Table I. Here, without loss of generality, $t_s$ for the Fe is defined as 0 fs. $t_s$ for Ni is hence the relative delay with respect to Fe. The amount of demagnetization ($m_d$) of Fe is ~ 27% more than that of Ni. The demagnetization time constant, $t_d$, for Fe is by ~ 29% larger than that for Ni, leaving the quenching rate, $m_d/t_d$ about the same for both the elements. A relative delay of 45 fs ± 9 fs in the Ni demagnetization is obtained from the fitting.

Table I: Results obtained from the fitting of the T-MOKE data presented in Fig. 1b.

|  | Fe | Ni |
|---|---|---|
| $m_d$ | $0.38 \pm 0.01$ | $0.30 \pm 0.01$ |
| $t_d$ (fs) | $221 \pm 13$ | $171 \pm 15$ |
| $t_s$ (fs) | $0 \pm 5$ | $45 \pm 7$ |

We now compare these results to the $L_3$ XMCD data measured at the Femtoslicing facility at the photon energy of the maximum XMCD effect at the Fe and Ni $L_3$ resonances, respectively. Here, unlike the T-MOKE setup, the element resolved magnetization cannot be measured simultaneously as the intrinsic bandwidth of the X-rays generated by the femto-slicing scheme is much smaller than the energy distance between the Fe and Ni $L_3$ resonances. Element specific information hence requires separate, sequential measurements at each elemental resonance. While the used femto-slicing setup has an intrinsic synchronization between pump and probe pulses, slow drifts may still occur due to, e.g., temperature changes affecting the beam paths. To ensure that the Fe and Ni dynamics is probed on identical delay scale, we used an X-ray-optical-cross-correlator (XOCC) as unique time reference for both measurements. This cross correlator records the laser-excited displacive coherent phonon oscillation of a MoSi multilayer structure, which can be probed at both resonance energies through the first super-lattice Bragg reflection in diffraction geometry (Fig. 2a). The time of laser excitation, has a defined reference to the phase of the displacive oscillation allowing to properly reference the delay scales of both $L_3$-edge measurements to each other. Detail of the XOCC setup can be found in Refs. [26–29] and are given in the supplement. We implemented a measurement protocol changing data acquisition between the sample and XOCC on an hourly base. The delay-axis for each XMCD scan was adjusted to the time reference as obtained from the MoSi XOCC data by assuming a linear drift of the delay scale in between (see supplement) .

The such obtained time-resolved XMCD data are shown in Fig. 2b. A delay of the Ni demagnetization compared to that of Fe is clearly visible. For quantitative analysis, we fitted the magnetization transients of Fe and Ni with the same function as for the T-MOKE data (see eqn. 2). The red and the blue curves are the fits to the data; the red and blue symbols are the data obtained by averaging over all recorded datasets. The fit results are summarized in Table II. Since no experimental time relation between the T-MOKE and the XMCD time axes is available, we here again define $t_s$ for Fe as 0 fs. For the delay of the Ni demagnetization onset with respect to that we find 42 fs with a standard error of 25 fs. This comes in almost perfect agreement with the results obtained from T-MOKE. Similar to the T-MOKE data, Fe demagnetizes (about 11%) more than Ni, while demagnetization times are similar for both elements. This contradicts with the findings of Ref. 7.

Table II: Results obtained from the fitting of the XMCD data presented in Fig. 2b.

|  | Fe | Ni |
|---|---|---|
| $m_d$ | $0.49 \pm 0.01$ | $0.44 \pm 0.01$ |
| $t_d$ (fs) | $264 \pm 28$ | $260 \pm 40$ |
| $t_s$ (fs) | $0 \pm 15$ | $42 \pm 20$ |

Discussion:

The experimental data of this study clearly confirm a delayed demagnetization for Ni in $Fe_{50}Ni_{50}$. Our results provide significant evidence that the delay found in the demagnetization onset of Ni relative to Fe in the $Ni_{50}Fe_{50}$ alloy is of intrinsic physical origin. Remarkably, we find even quantitative agreement for the relative-delay: $(42 \pm 25)$ fs in XMCD and $(45 \pm 9)$ fs in T-MOKE. Given non-linearity and cross-talk in T-MOKE such a good agreement seems rather surprising.

As for the cross-talk, we observe that the maximum asymmetries for Fe and Ni (Fig. 1b) are of similar magnitude. Their magnitude differs from the asymmetries of the elemental metals (see e.g. Ref. 22): The much larger asymmetry of elemental Ni would lead to a rather strong Ni contribution at the Fe peak energy. In the particular alloy studied here, this cross-talk effect will clearly be weaker. Other contributions to non-linearities show quite pronounced energy dependences (also seen dynamically in Ref. 23). Since the probing energy of HHG sources is limited to integer multiples of the fundamental energy, an overall linear response in T-MOKE must be seen as somewhat fortuitous. The deviation in the decay times ($t_d$) and demagnetization amplitude ($m_d$) between the two experiments can be assigned to different pump fluence levels. Although we measured shorter decay times for T-MOKE compared to XMCD, the quenching rate ($m_d/t_d$) turns out to be very similar within the error bars. The latter is in agreement with findings of Ref. [30].

Conclusions:

Our comparative XMCD and T-MOKE study shows that a delay in the Ni demagnetization in $Fe_{50}Ni_{50}$ is an intrinsic effect and not caused by non-linearity or cross-talk issues in the way T-MOKE is probing magnetization. The amount of delay observed with both techniques is found to agree remarkably well. With both the methods, the amount of demagnetization of the Fe was found to be somewhat larger compared to the Ni. The demagnetization time ($t_d$) and the amount of demagnetization ($m_d$) obtained from the XMCD measurements are larger in general compared to that of the T-MOKE measurements, while the quenching rate, $m_d/t_d$ comes out about the same for both the techniques. While in this study, T-MOKE results appear reasonably correct, this does not imply that T-MOKE should not be treated with caution – in particular when the involved resonances are close in energy.

By reliably evidencing the hitherto controversially discussed delayed demagnetization of Ni in FeNi alloy, this experimental study supports the validity of existing theory work and ensures further progress on the field of magnetic dynamics in technologically important multi-element materials.

Acknowledgements:

We acknowledge instructive discussions with Sangeeta Sharma. This work received funding from the DFG within the Transregio TRR227 Ultrafast Spin Dynamics (Project A03).

# Supplementary Materials: Experimental confirmation of the delayed Ni demagnetization in FeNi alloy

Fitting of the displacive oscillation in MoSi-multilayer:

The physics of the displacive coherent phonon oscillation (DCPO) is described in Ref. 26, 27. In brief, the excitation of a [Mo(1.86 nm)/Si(2.07 nm)]$_{40}$ multilayer by an infrared fs-laser pulse leads to preferential absorption by the metallic Mo layers while the Si layers essentially remain unexcited. The optical absorption and subsequent thermalization induces an increased equilibrium average distance of the Mo atoms towards which they begin to relax. This displacive excitation leads to a subsequent damped expansion / compression oscillation of all Mo layers around a new equilibrium layer width.

X-rays can probe the MoSi superstructure via the corresponding Bragg superlattice peaks. The laser induced coherent displacive oscillation translates into an oscillatory change of the Bragg peak amplitude, which is independent of the probing X-ray wavelength. The DCPO, or more precisely its phase, can therefore be used as a unique time reference for the pump-probe experiment. We measured the laser induced oscillation of the first order super-lattice peak (SL1) amplitude at both Fe and Ni $L_3$ edge photon energies, respectively, alternating with the corresponding XMCD measurements on the alloy sample. The obtained data are plotted in Fig. 2a of the main manuscript.

The transient change of peak height and subsequent oscillation around the new equilibrium, characteristic of a displacive oscillation is described by

$$C_{DCPO}(t) = G(t) \otimes R_{DCPO}(t)$$

Here, *G(t)* is a Gaussian function with FWHM = 130 fs to take into account the overall temporal resolution, which is convoluted by the normalized response function $R_{DCPO}(t)$ of the DCPO. The latter is assumed as

$$R_{DCPO}(t) = 1, \qquad \text{for } t \leq 0$$

$$R_{DCPO}(t) = -A \left\{ 1 - exp\left(-\frac{t}{T}\right) \cdot \cos\frac{2\pi}{p}(t - t_\varphi) \right\} + m \cdot t, \qquad \text{for } t > 0,$$

where the constant *A* represents the amplitude of the displacement leading to an intensity drop of the SL peak intensity upon laser excitation. The second term inside the curly brackets represents the damped oscillation with the amplitude A, where *T* is the damping time of the oscillation, *p* the oscillation period, and $t_\varphi$ the phase offset of the cosine function with respect to time-zero. The latter entails the delay of the oscillation onset. The linear slope *m* accounts for the shift of the Bragg reflection due to the average heat expansion of the superlattice [28]. For high fluences as used in our experiment, the DCPO starts without delay upon laser excitation, yielding $t_\varphi = 0$. [29] The fit results of $C_{DCPO}(t)$ to the experimental data are shown as dashed lines in Fig. 2a of the main manuscript. The fitted results at both Fe and Ni $L_3$ edges are summarized in the Table S1.

Table S1: Results obtained from the fitting of the oscillation of SL1 peak intensity in MoSi at Fe and Ni L$_3$ edges.

|  | 707 eV (Fe) | 840 eV (Ni) |
|---|---|---|
| $A$ (%) | 0.84 ± 0.07 | 0.65 ± 0.04 |
| $p$ (fs) | 588 ± 21 | 588 ± 15 |
| $t_\varphi$ (fs) | 0 ± 15 | 0 ± 10 |